\documentclass[prl,
 reprint,
showpacs,preprintnumbers,
 amsmath,amssymb,
 aps,
]{revtex4-1}
\usepackage{graphicx}
\usepackage[pdftex]{hyperref}
\usepackage[T1]{fontenc}
\usepackage[latin1]{inputenc}
\begin{document}

\title{Large spiral and target waves: Turbulent diffusion boosts scales of pattern formation}

\author{A. von Kameke}\email[Corresponding author:\,]{alexandra.vonkameke@usc.es}
\author{F. Huhn}
\author{A.P. Mu\~nuzuri}
\author{V. P\'erez-Mu\~nuzuri}

\affiliation{Group of Nonlinear
Physics, University of Santiago de Compostela. E-$15782$ Santiago de
Compostela, Spain}

\date{September 2012}

\begin{abstract}
In absence of advection, reaction-diffusion systems are able to organize into spatiotemporal patterns, in particular spiral and target waves. Whenever advection is present and can be parameterised in terms of effective or turbulent diffusion $D_{*}$, these patterns should be attainable on much greater, boosted lengthscale. However, so far, experimental evidence of these boosted patterns in turbulent flow was lacking. Here, we report the first experimental observation of boosted target and spiral patterns in an excitable chemical reaction in a quasi two-dimensional turbulent flow. The wave patterns observed are $\sim 50$ times larger than in the case of molecular diffusion only. We vary the turbulent diffusion coefficient $D_{*}$ of the flow and find that the fundamental Fisher-Kolmogorov-Petrovsky-Piskunov (FKPP) equation $v_{f} \propto \sqrt{D_{*}}$ for the asymptotic speed of a reactive wave remains valid. 
However, not all measures of the boosted wave scale with $D_{*}$ as expected from molecular diffusion, since the wavefronts turn out to be highly filamentous. 
\end{abstract}

\pacs{}

\maketitle
Pattern formation in reaction-diffusion-advection (RDA) systems is an important process in many natural and man-made systems, e.g., plankton growth and iron fertilization in the ocean\,\cite{smetacek2012}, dispersion of pollutants in the atmosphere, and optimal mixing in chemical reactors\,\cite{neufeld2004}. 
Spiral and target waves have been observed on small scales in various active media, e.g. in chicken retina\,\cite{yu2012}, cardiac tissue\,\cite{luther2011} or chemical reactions\,\cite{perezmunuzuri1991, kapral1995}. From a geophysical viewpoint it is of crucial interest if these reaction-diffusion patterns can also be found in large scale systems involving turbulent advection, as for example plankton dynamics in the ocean affecting CO$_{2}$ absorption\,\cite{smetacek2012,neufeld2004,huhn2012}. 
Theoretically, the appearance of spiral and target waves should be possible in RDA systems whenever the advection term can be parameterised as a global diffusion coefficient\,\cite{grigoriev2011}. However, so far, experimental evidence of these patterns in turbulent flows is lacking.
Despite the importance of pattern formation in RDA systems only very few laboratory experiments on turbulent fluid flow involve reaction kinetics\,\cite{ronney1995}, and to our knowledge, none has considered excitable kinetics so far. Considerable numerical and experimental effort has focused on cellular and chaotic flows due to the simpler realization\,\cite{boehmer2008,arratia2006,neufeld2004,majda1999,boehmer2008}.
In this Letter, we show experimentally that pattern formation, in particular, spiral and target waves can occur in turbulent fluid flows and we find that the front expansion is limited by the FKPP equation.

We create a quasi two-dimensional turbulent flow using the Faraday experiment\,\cite{kameke11,haslam1995}, i.e. we vertically vibrate a circular container of $30$\,cm diameter filled with $2$\,mm of an excitable cyclohexandione and ferroin based Belousov-Zhabotinsky reaction (BZ)\,\cite{kameke10} (see methods summary and supplementary Fig.\,S1\,\cite{supps}). The dynamics of this chemical reaction can be well observed in the visible range due to the oxidation of the reddish catalyst ferroin $[Fe(phen)_{3}^{2+}]$ to the blue ferriin $[Fe(phen)_{3}^{3+}]$ \,\cite{kurin1996}. We vary the intensity of the turbulence and thus the turbulent diffusion constant\,\cite{majda1999} $D_{*}$ by altering the amplitude $a_{0}$ of the acceleration and the frequency $f$ of the vertical forcing. 
%
\begin{figure*}[htb!]
\includegraphics[width=\textwidth]{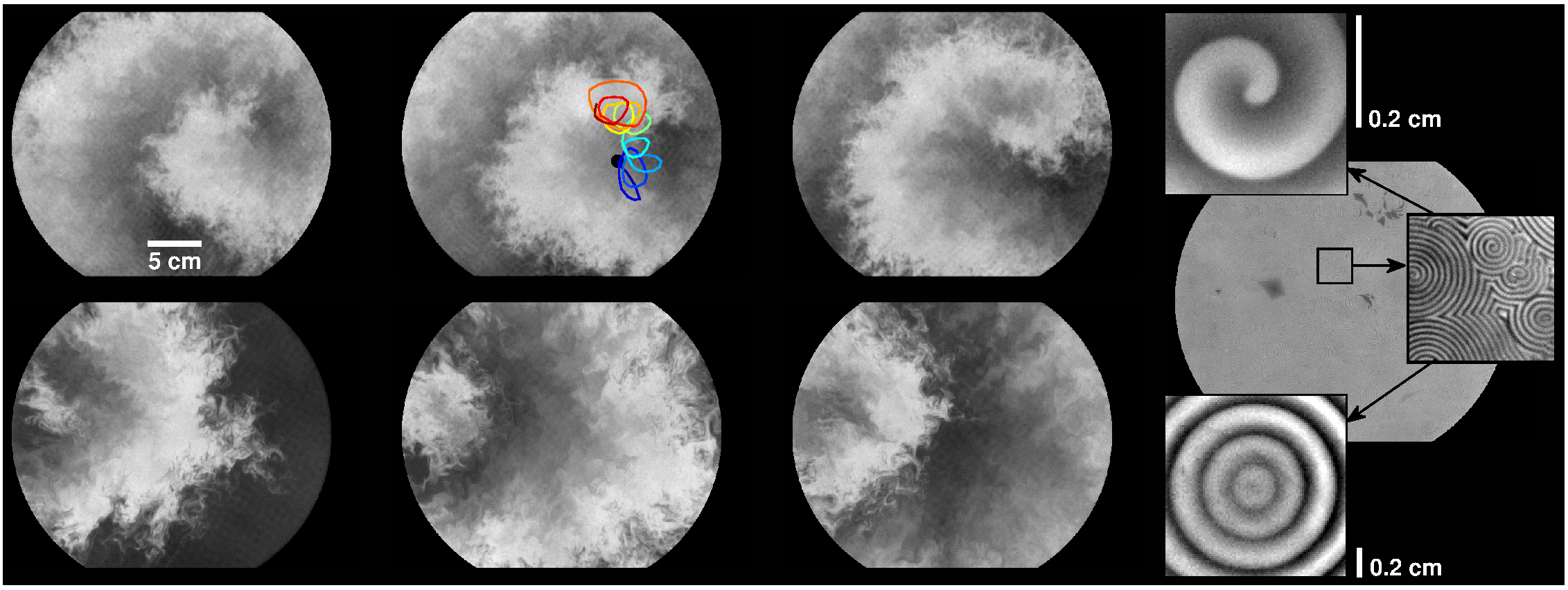}
\caption{(Color online) Boosted spiral and target patterns in a turbulent flow.
Greyscale indicates concentration of ferriin, $Fe(phen)_{3}^{3+}$.  Upper panel, $1$-$3$: Image sequence of boosted spiral with $\Delta t \approx 3.6$\,s, $f = 70$\,Hz, $a_{0} \approx 1.8$\,$g$, $[H_{2}SO_{4}] = 1.2$\,M, period of spiral $T = (13\pm1)$\,s. $2$nd image: Trajectory of spiral tip in time (color code:  early position blue, late position red). Lower panel, $1$-$3$: Image sequence of boosted target wave with $\Delta t \approx 12.4$\,s, $f = 50$\,Hz, $a_{0} \approx 1.2$\,$g$, $[H_{2}SO_{4}] = 0.6$\,M. Both patterns form spontaneously and are persistent phenomena that can last from a few minutes up to one hour. 
For corresponding movies  (M1, M2) see supplementary material\,\cite{supps}. Right: Three close-ups show molecular-diffusion-induced spiral and target patterns in absence of fluid flow in the same container. Note the large difference in scales between these usual and the boosted patterns. }
\label{fig:BigSpiral}
\end{figure*}

Figure\,\ref{fig:BigSpiral} shows examples of the boosted patterns in the turbulent flow. The upper panel, $1$--$3$, shows an image sequence of a spontaneous boosted spiral and the lower panel, $1$--$3$, a spontaneous boosted target wave (supplementary movies M1 and M2\,\cite{supps}). Without any fluid flow the much smaller usual target and spiral patterns can be observed which are shown for comparison on the right (image $4$). The boosted patterns are a very robust phenomenon and were found for a large range of forcing parameters, $f = 30$--$140$\,Hz, $a_{0} = 0.6$--$2.5$\,$g$, $g$ being the gravitational constant. The temporal persistence of the target patterns varies from some minutes for high forcing amplitudes $a_{0}$, to up to one hour for lower ones. The probability for a target to form is higher for lower forcing. This is most likely related to more long-lived structures in the fluid flow\,\cite{perezmunuzuri2010} that favor the occurrence of a perturbation that is persistent and big enough to trigger a new wave\,\cite{foerster1989}. Usually, but not always, target waves are triggered at the border of the container. Spiral waves form spontaneously, most often created by the breakup of target waves due to interactions with the turbulent fluid flow or the boundary, but they can also be created intentionally by an abrupt short interruption of the forcing. Figure\,\ref{fig:BigSpiral} (upper panel, image $2$) shows the trajectory of a spiral tip in time. The temporal persistence of the spiral is limited due to the complex movement of the tip \,\cite{mikhailov1990} since it eventually hits the border or another pattern, causing the spiral to vanish (supplementary movies M1 and M3\,\cite{supps}). Qualitative observations suggest that the displacement of the spiral tip is a superposition of a random movement due to the filamentary structure of the front and a migration along the border of the container \,\cite{mikhailov1995}.

For a quantitative analysis of the periods of the boosted spirals we varied the turbulent diffusion of the flow. This was achieved by changing only the forcing amplitude $a_{0}$ leaving the forcing frequency, and thus the Faraday wavelength $\lambda_{F}$ constant\,\cite{haslam1995,tufillaro1989} ($f = 50$\,Hz, $[H_{2}SO_{4}] = 0.6$\,M, supplementary example movie M3\,\cite{supps}). The periods of the boosted spirals at $f= 50$\,Hz are in the range $T = 30$--$50$\,s for all forcing amplitudes with a slight tendency towards higher periods for stronger forcings. This might be explained by the augmentation of the width of the boosted autowaves such that the spirals seem to be restricted by their own tail\,\cite{mikhailov1990}. This self-restriction could also explain why the period of the molecular-diffusion-induced spiral, $T_{mol} = 18$--$25$\,s, was somewhat lower. Further, in order to prevent the spiral to drift, we pinned its tip to a round obstacle of $54$\,mm diameter, placed in the middle of the container. These pinned spirals last for up to $\sim 1$\,h (see supplementary data, movies M4, M5 and M6\,\cite{supps}). 
In addition to the spiral and target patterns we also observe double spirals with two free curling ends (supplementary movie M7\,\cite{supps}), as well as up to $3$ simultaneously existing spirals. All reactive waves had the typical characteristics of autowaves, in particular, they annihilate when they meet. 
%
\begin{figure*}[htb!]
\includegraphics[width=\textwidth]{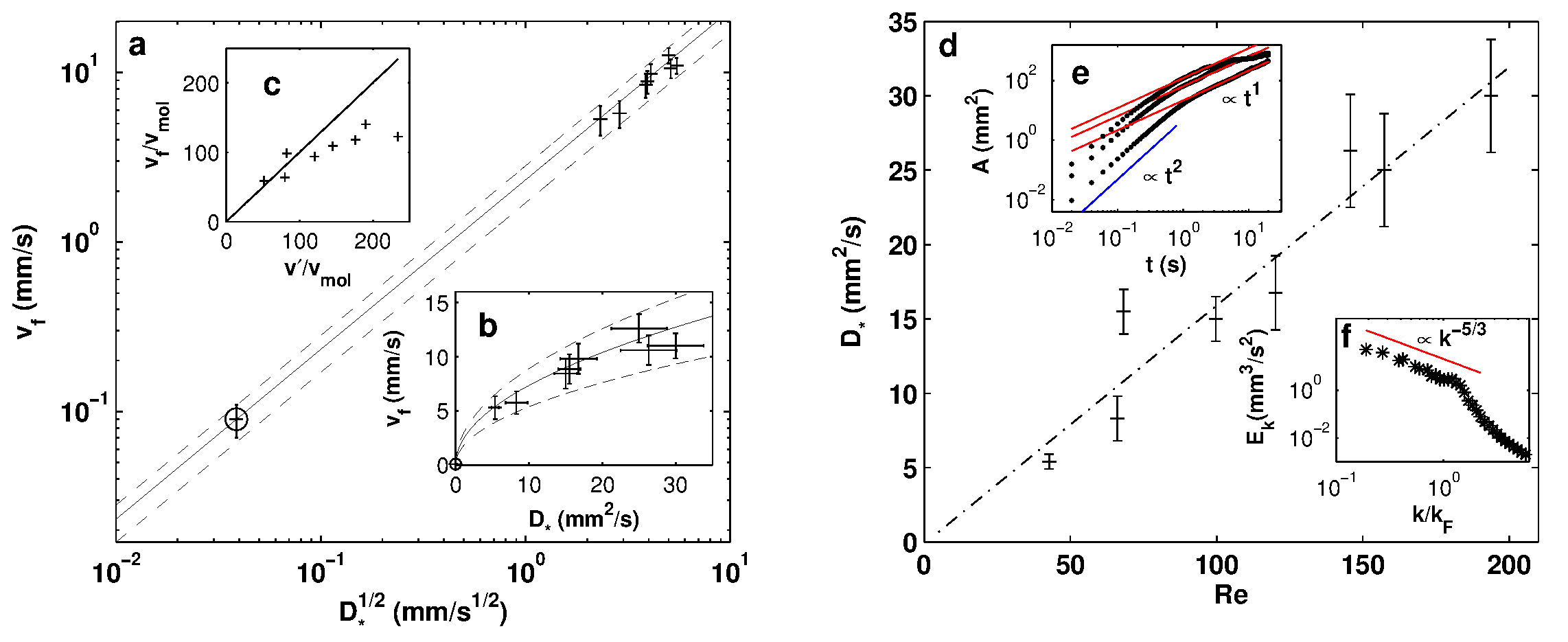}
\caption{(Color online) Front velocity of reactive waves in dependence of turbulent diffusion. 
(a) The velocity of the target wave fronts $v_{f}$ scales with $\sqrt{D_{*}}$ and follows the FKPP prediction $v_{f} = 2\sqrt{D/\tau_{reac}}$ (solid line). The time constant of the reaction $\tau_{reac} = (0.8 \pm 0.3)$\,s was derived from the molecular case (circle) but adjusts also well for the turbulent data (crosses). Dashed lines indicate the error bounds estimated from the standard deviation of the velocity measurements from the molecular-diffusion-induced target wave. Inset (b) shows a close up of the turbulent data pairs. (c) Target front velocity $v_{f}$ vs. turbulent root-mean-square velocity in one direction $v' = v_{rms}/\sqrt{2} $, both normalized to the front velocity $v_{mol}$ of the molecular-diffusion-induced target wave. (d) The measured diffusion coefficients are shown as a function of the Reynolds number $Re = v_{rms} \lambda_{F}/ \nu$ indicating the turbulence strength, where $\nu$ is the kinematic viscosity of the fluid. Inset (e) shows exemplary the absolute diffusion for the flows with $Re \approx 43$, $Re \approx 120$ and $Re \approx 194$ and the linear fit for estimation of the turbulent diffusion coefficient. Inset (f) shows an exemplary energy spectra of the flow for $Re \approx 120$. A double cascade and a regime with a Kolmogorov type scaling ($ E_{k} \propto k^{-5/3}$) can be distinguished. $k_{F}$ is the typical Faraday wavenumber.}
\label{fig:vddRe}
\end{figure*}

Figure\,\ref{fig:vddRe}\,(a) and inset (b) show that the FKPP relation for the front velocity $v_{f}$ remains valid for well developed boosted target waves in the quasi two-dimensional turbulent flow, i.e., $v_{f} = 2\sqrt{D_{*}/\tau_{reac}}$, $\tau_{reac}$ being the reaction timescale\,\cite{grigoriev2011}. Surprisingly, the boosted data points agree with the prediction derived from the FKPP equation using only measurements from experiments with molecular diffusion: The solid line is the solution of the FKPP relation, where the typical reaction timescale $\tau_{reac}$ was estimated from the velocity measurement of the molecular-diffusion-induced target wave to be $\tau_{reac} = (0.8 \pm 0.3)$\,s and the molecular diffusion coefficient was estimated from the literature to be $D_{mol} \approx (1.3\: \text{-} \: 2.0) \: 10^{-3} \: \textrm{mm} ^{2} / \textrm{s} $\,\cite{mori1991,kuhnert1985,foerster1989}. 
Theoretically, when the reaction timescale is small in comparison to the timescale of the fluid flow, the front velocity $v_{f}$ is bounded by the unidirectional root-mean-square velocity of the flow instead of obeying the FKPP relation\,\cite{brandenburg2011}. Inset (c) shows that in our experiments this limit is only approached for low forcing. 
We noted that the variation of the front velocity is related to the interval in between successive waves which suggests that they might obey a dispersion relation analogue to usual target waves\,\cite{dockery1988}.
 
In Fig.\,\ref{fig:vddRe}\,(e) the measured turbulent diffusion coefficient $D_{*}$ is plotted as a function of the estimated Reynolds number for different forcing amplitudes. The turbulent diffusion increases approximately linearly with the Reynolds number as expected, and mixing is enhanced. At these Reynolds numbers the flow is turbulent as can be seen in an exemplary energy spectrum ($Re \approx 120 $) revealing a double cascade and a Kolmogorov type scaling ($\propto k^{-5/3}$) in inset Fig.\,\ref{fig:vddRe}\,(f)\,\cite{kameke11,ronney1995,boffetta2012}. The turbulent diffusion coefficients $D_{*}$ were estimated from measurements of the absolute dispersion $A(t)$ shown in Fig.\,\ref{fig:vddRe} inset (e), by a fit to the regime of linear growth.
\begin{figure}[htb!]
\includegraphics[width=8.5cm]{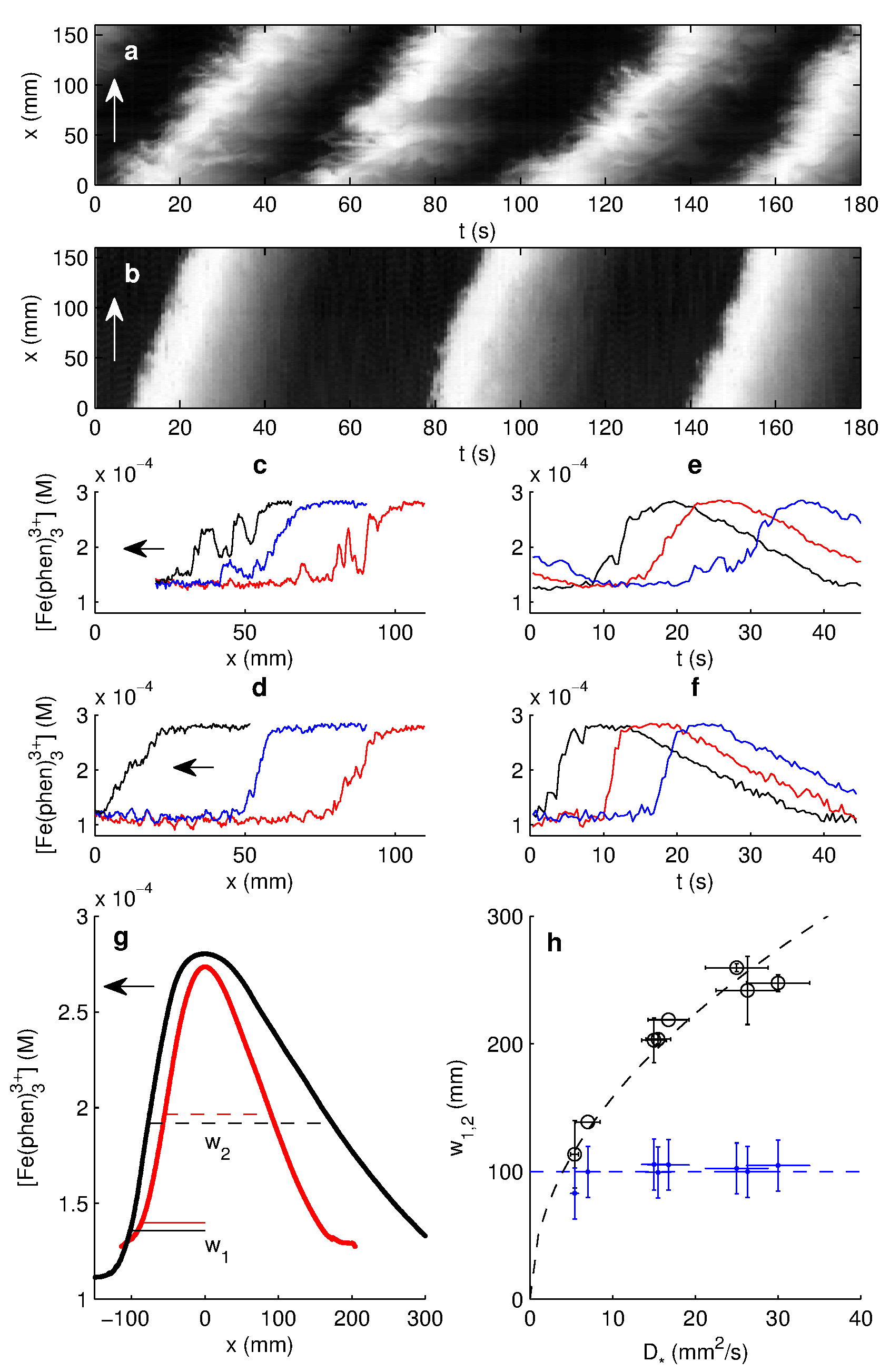}
\caption{(Color online) Front characteristics of boosted target waves. (a),\,(b) Space-time plots of boosted targets for $D_{*} \approx 5.4\: \textrm{mm} ^{2} / \textrm{s} $ $(a \approx 1.3\,g_{0})$ and $D_{*} \approx 30.0\, \textrm{mm} ^{2} / \textrm{s} $ $(a \approx 2.2\,g_{0})$, arrows indicate the direction of front propagation (supplementary movies M2 and M8\,\cite{supps}). The target waves are narrower, slower and more filamentous for the smaller diffusion coefficient. (c),\,(d) Ferriin concentration, [$Fe(phen)_{3}^{3+}$], along a line at three different instances of time, $\Delta t \approx 6.4$\,s, for $D_{*} \approx 5.4\: \textrm{mm} ^{2} / \textrm{s} $ and $D_{*} \approx 30.0\: \textrm{mm} ^{2} / \textrm{s} $ respectively. (e),\,(f) Ferriin concentrations for the same values of $D_{*}$ at three different points in space ($\Delta x \approx 53$\,mm, $\Delta x \approx 80$\,mm). (g) The mean profile of the target waves for both diffusion coefficients estimated by averaging over all targets measured. (h) Different widths $w_{1}$ and $w_{2}$ of the profile in dependence of the diffusion coefficient $D_{*}$. The full width $w_{2}$ of the target wave grows with $\sqrt{D_{*}}$ as expected while the width of the rising edge $w_{1}$ stays constant.}\label{fig:sptplnwidth}
\end{figure}

Despite the validity of the FKPP prediction for the front speed, Fig.\,\ref{fig:sptplnwidth} demonstrates that the boosted target waves do not entirely behave like their molecular diffusion counterparts. An important difference is the complex filamentous structure of the reaction front which is related to the small scale stretching and folding processes in the turbulent dynamics (Fig.\,\ref{fig:sptplnwidth}\,(a),\,(b) and Fig.\,\ref{fig:BigSpiral})\,\cite{koudella2004,brandenburg2011,ronney1995}. 
For smaller turbulent diffusion (Fig.\,\ref{fig:sptplnwidth}\,(a)) the filamentary structure increases due to two distinct processes: 
First, the increase of the length and persistence of the filaments can be explained by coherent flow structures, i.e. little eddies and jets, that order the flow on timescales longer than the reaction time $\tau_{reac}$. An imprint of the filaments can be seen in the ferriin concentration profiles (Fig.\,\ref{fig:sptplnwidth}\,(c)). The peaks of high concentration ahead of the front show the intermittency of the turbulent diffusion process on these spatiotemporal scales. For higher turbulent forcing the fronts are less intermittent (Fig.\,\ref{fig:sptplnwidth}\,(d)).
Second, the sharper and more pronounced appearance of the filaments can be explained by the Damköhler number, $Da = \tau_{flow}/\tau_{reac}$, the ratio of the typical timescales of the flow and the reaction. The flow timescales were estimated to be the ratio of the Faraday wavelength and the root-mean-square flow velocity, $\tau_{flow} = \lambda_{f}/v_{rms}$. $Da$ varied from $Da \approx 0.4$ for the highest forcing to $Da \approx 1.8$ for the lowest (supplementary Fig.\,S2\,\cite{supps}). 
For small $Da$, the fluid flow is fast compared to the reaction timescale which causes the front to be smoother in agreement to what we find for strong forcing. 
For large $Da$, and thus lower forcing, the front appears sharper and its velocity approaches the root mean square velocity in one direction, $v' = v_{rms}/\sqrt{2}$\,\cite{koudella2004,brandenburg2011}. This limit is reached in our experiments for small forcings as is reflected by inset (c) in Fig.\,\ref{fig:vddRe}. 

Furthermore, in Fig.\,\ref{fig:sptplnwidth}\,(a),(b), it is easy to observe by eye the differences in the target front velocities, the frequencies of spontaneous target formation and the target widths for the two extreme cases of the measured turbulent diffusion. In order to quantify the dependence of the width on the turbulent diffusion, Fig.\,\ref{fig:sptplnwidth}\,(g) depicts the mean profiles of the boosted target for the two turbulent diffusion coefficients. These measurements were repeated for all turbulent diffusion coefficients (Fig.\,\ref{fig:sptplnwidth}\,(h)). While the full width $w_{2}$ of the boosted target waves increases according to $w_{2} \propto \sqrt{D_{*}}$, as expected for an ideal reaction-diffusion system\,\cite{murray1989}, the width of the rising edge $w_{1}$ does not change within the error of the measurement. 
A possible explanation for this unexpected behavior of $w_{1}$ is the intermittency of the mixing process: Averaging over many sharply defined filaments could give a similar width for the mean profile as the average over a smoother and broader front. This indicates that for low forcings and on the timescales of the fast forward reaction occurring at the leading edge of the front mixing might not yet be well defined by a diffusive process.
According to this picture, $w_{2}$ augments diffusively as the backward reaction at the tail of the front is much slower and sees a well developed diffusive process.
Timescales of the forward and the backward reaction can be estimated as the times of rise and fall of the ferriin concentration in Fig.\,\ref{fig:sptplnwidth}\,(e),\,(f).

In summary, we conclude that complex spatiotemporal patterns, such as target and spiral waves, occur in turbulent fluid flows as was shown experimentally. Measuring turbulent diffusion coefficients and the reaction front velocities at various Reynolds numbers we find that they obey the FKPP relation for reaction-diffusion systems. The overall patterns resemble those of their molecular counterparts, however, an important difference is the filamentary appearance of the front which leads to an unexpected scaling of the front width. We suggest that this phenomena can be understood by the existence or absence of coherent structures in the flow that are known to exist in many turbulent flows. 
We expect our results to increase the attention on pattern formation in systems where excitable dynamics evolve in turbulent flows, such as plankton growth in the ocean where a ring-like structure, similar to a target, has been reported\,\cite{wyatt1973,dubois1975}.  

\begin{acknowledgements}
This work was supported by the Ministerio de Educacion y Ciencia under Research Grants No. FIS$2010-21023$.
A.v.K. and F.H. receive funding from FPU, No. AP-2009-0713 and AP-2009-3550. 
\end{acknowledgements}


\end{document}